\documentclass[twocolumn,pra,aps,showpacs]{revtex4}

\usepackage{amsmath}
\usepackage{amssymb}
\usepackage[dvips]{graphicx}
\usepackage{pstricks}
\usepackage{pst-node}
\usepackage{pst-plot}

\newcommand{\qed}{\hspace*{\fill}$\square$}

 \newcommand{\R}{\mathbf{R}}

 \newcommand{\Z}{\mathbf{Z}}

 \newcommand{\paratodo}{\forall\,}

 \newcommand{\sset}[1]{ \{#1\} }
 \newcommand{\half}{\frac 1 2}

 \newcommand{\ket}[1]{|#1\rangle}

\begin{document}

\title[Short Title]{
Exact Topological Quantum Order in $D=3$ and Beyond: \\Branyons and
Brane-Net Condensates}

\author{H. Bombin and M.A. Martin-Delgado}
\affiliation{
Departamento de F\'{\i}sica Te\'orica I, Universidad Complutense,
28040. Madrid, Spain.
}

\begin{abstract}
We construct an exactly solvable Hamiltonian acting on a
3-dimensional lattice of spin-$\frac 1 2$ systems that exhibits
topological quantum order. The ground state is a string-net and a
membrane-net condensate. Excitations appear in the form of
quasiparticles and fluxes, as the boundaries of strings and
membranes, respectively. The degeneracy of the ground state depends
upon the homology of the 3-manifold. We generalize the system to
$D\geq 4$, were different topological phases may occur.
The whole construction is based on certain special complexes that we
call colexes.
\end{abstract}


\maketitle

\section{Introduction}
\label{intro}

Deviations from a standard theory in a certain field of Physics
has always attracted the attention of searching for new physics. In
condensed matter, the standard model is the Landau Theory of
quantum liquids (Fermi liquid) supplemented with the Spontaneous
Symmetry Breaking mechanism (SBB) and the Renormalization Group
scheme \cite{landau37}, \cite{ginzburg_landau50},
\cite{wenbook04}. The concept of local order parameter plays a central
role in detecting quantum phases or orders within Landau's theory.
Quite on the contrary, topological orders cannot be described by means of local order parameters
or long range interactions. Instead, a new set of quantum numbers is needed
for this new type of phases, such as ground state degeneracy, quasiparticle braiding statistics,
edge states \cite{wenniu90}, \cite{wen90}, \cite{wen92},
topological entropy  \cite{kitaevpreskill06}, \cite{levinwen06}, etc.

A consequence of the SBB is the existence of a ground state degeneracy.
However, in a topological order there exists ground state degeneracy
with no breaking of any symmetry. This degeneracy has a topological origin.
Thus, topological orders deviate significantly from more standard orders covered
within the Landau symmetry-breaking theory.
The existence of topological orders seems to indicate that
nature is much richer than the standard theory has predicted so far.

Emblematic examples of topological orders are  Fractional Quantum Hall Liquids (FQH).
FQH systems contain many different phases at T=0
which have the same symmetry. Thus, those phases cannot be distinguished
by symmetries and Landau's SBB does not apply
\cite{wenniu90}, \cite{blokwen90}, \cite{read90}, \cite{frohlichkerler91}. Then, we need to resort
to other types of quantum numbers to characterize FQH liquids. For example, the
ground state degeneracy $d_g$ depends on the genus $g$ of the $D=2$ surface
where the electron system is quantized, namely, $d_g = m^g$ with
 filling factor being $\nu = \frac{1}{m}$.

There are several other examples of topological orders like
short range RVB (Resonating Valence Bond) models \cite{roksharkivelson88}, \cite{readchakraborty89},
\cite{moessnersondhi01}, \cite{ardonne_etal04},
quantum spin liquids \cite{kalmeyerlaughlin87}, \cite{wenwilczekzee89}, \cite{wen90}, \cite{readsachdev91},
\cite{wen91}, \cite{senthilfisher00}, \cite{wen02}, \cite{sachdevparks02}, \cite{balentsfishergirvin02}
etc. Due to this topological
order, these states exhibit remarkable entanglement properties
\cite{martindelgado04}, \cite{martindelgado04b}.
Besides these physical realizations, there have been other proposals for
implementing topological orders with optical lattices \cite{duandemlerlukin03}, \cite{zoller05}, \cite{pachos05}
with spin interactions in honeycomb lattices \cite{kitaev05}.
In this paper we shall be concerned with topological models
constructed with spins $S=\half$ located at the sites of certain
lattices with coordination number, or valence, depending on the dimension $D$
of the space and the property of being coloreable to be explained in Sect.\ref{seccionII}.

From the point of view of quantum information \cite{rmp}, a
topological order is a new type of entanglement: it exhibits
non-local quantum correlations in quantum states. A topological
phase transition is a change between quantum states with different
topological orders. In dimensions $D \geq 4$ we construct exact
examples of quantum lattice Hamiltonians exhibiting  topological
phase transitions in Sect.\ref{seccionIVA}. Here we find an example
of topology-changing transition  as certain coupling constant is
varied in $D=4$. This is rather remarkable since the most usual
situation is to have an isolated topological point or phase
surrounded by non-topological phases \cite{martindelgado04},
\cite{martindelgado04b}.

In two dimensions, a large class of  ``doubled" topological phases has been
described and classified mathematically using the theory of tensor categories \cite{levinwen05}.
The physical mechanism underlying this large class of topological
orders is called string-net condensation. This mechanism is the equivalent mechanism
to particle condensation in the emergence of ordered phases in Landau's theory.
A string-net is a network of strings and it is a concept more general than a collection
of strings, either closed or open. In a string-net we may have the situation in which a
set of strings meet at a branching point or node,
something that is missing in ordinary strings which have two ends at most (see Fig.\ref{figura_particulas}).
More specifically, the ground state of these theories are described by superpositions
of states representing string-nets. The physical reason for this is the fact that local energy constraints
can cause the local microscopic degrees of freedom present in the Hamiltonian to organize into
effective extended objects like string-nets.

A new field of applications for topological orders has emerged with the theory
of quantum information and computation
\cite{kitaev97}, \cite{freedman98}, \cite{dennis_etal02}, \cite{bravyikitaev98}.
Quantum computation, in a nutshell, is the
art of mastering quantum phases to encode and process information. However, phases
of quantum states are very fragile and decohere. A natural way to protect them from
decoherence is to use topologically ordered quantum states which have a non-local kind of entanglement.
The non-locality means that the quantum entanglement is distributed among many different particles in
such a way that it cannot be destroyed by local perturbations. This reduces decoherence  significantly.
Moreover, the quantum information encoded in the topological states can be manipulated by moving quasiparticle
excitations around one another producing braiding effects that translate into universal quantum gates \cite{kitaev97},
\cite{Ogburn99}, \cite{freedman_etal00a}, \cite{freedman_etal00b}, \cite{freedman_etal01}, \cite{preskillnotes}.
Nevertheless, there are also alternative schemes to do lots of quantum information tasks by only using
the entanglement properties of the ground state \cite{topologicalclifford}, \cite{homologicalerror}, \cite{optimalGraphs}.

The situation for topological orders in $D=3$ is less understood. This is in part due to the very intrincate
mathematical structure of topology in three dimensions. While the classification of all different topologies
is well stablished in two dimensions, in $D=3$ the classification is much more difficult and only recently it
appears to be settled with the proof of the Thurston's geometrization conjecture \cite{thurston},
a result that includes the Poincar\'e conjecture as a particular case \cite{perelmanI}, \cite{perelmanII},
\cite{perelmanIII}.
Topological orders have been investigated in three dimensions with models that exhibit string-net condensation
\cite{levinwen05} using trivalent lattices that extend the case of trivalent lattices in two dimensions.
However, a problem arises when one wishes to have an exactly solvable Hamiltonian describing this topological
phase since this type of magnetic flux operators do not commute in three dimensions any more.
A solution to this problem
can be found by imposing additional constrains to the mechanism found in $D=2$, but this obscures somehow
the geometrical picture of the resulting exactly solvable model.
Alternatively, it is possible to use a 3D-generalization
of Kitaev's toric code to provide examples with topological order based both in string condensation and
membrane condensation \cite{hammazanardiwen05}. In the theory of topological quantum error correcting codes,
there are also studies of toric codes in dimensions higher than $D=2$
\cite{dennis_etal02}, \cite{wang_etal03}, \cite{takeda04}.

In this paper we introduce a new class of exactly solvable models in $D=3$ that exhibits topological order.
Here we construct a class of models in which magnetic flux operators of several kinds commute among each other.
This is achieved by requiring certain geometrical properties to the lattices where the models are defined on.
As a result, we can study the whole spectrum of the models and in particular their quantum topological
properties. The ground state can be described as a string-net condensate or alternatively, as a membrane-net
condensate. A membrane-net condensate is a generalization of a collection of membranes, much like string-nets
generalize the notion of strings. Thus, in a membrabrane-net,
membranes can meet at branching lines instead of points.
Excitations come into two classes: there are quasiparticles that appear as the end-points of strings,
or certain type of fluxes that appear as the boundaries of membranes. These fluxes are extended objects.
Interestingly enough, when a quasiparticle
winds around a closed flux, the system picks up a non-trivial Abelian phase (see Fig.\ref{figura_winding}),
much similar like when one anyon \cite{wilczek82}, \cite{leinaasmyrheim77}
winds around another anyon acquiring an Abelian factor in the wave function of the system.
We coin the name branyons to refer to this quasiparticles that are anyons with an extended structure.
In fact, in our models they appear as Abelian branyons.

Our constructions can be nicely generalized to higher dimensions and we can compute exactly the ground
state degeneracies in terms of the Betti numbers of the manifolds where the lattice models are defined.
This allows us to discriminate between manifolds with differente homological properties using quantum
Hamiltonians. The generalized membranes are called branes and we find also a brane-net mechanism.

This paper is organized as follows: in Sect.\ref{seccionII} we introduce the models defined in three
dimensional lattices placed on different manifolds. These lattices are constructed by means of color
complexes that we call colexes of dimension 3, or 3-colexes; in Sect.\ref{seccionIII} the notion of
colexes is generalized to arbitrary dimensions; in Sect.\ref{seccionIV} we extend the topological
quantum Hamiltonians beyond $D=3$ dimensions and in particular we find instances of topology-changing
phase transitions; Sect.\ref{seccionV} is devoted to conclusions. In a set of appendices we provide
a full account of technical details pertaining to particular aspects of our models.

\section{The model in 3-manifolds}
\label{seccionII}

\subsection{Topological order and homology}
\label{seccionIIA}

The model that we are going to study belongs to the category of
topologically ordered quantum systems. A system with topological
quantum order is a gapped system that shows a dependency between the
degeneracy of its ground state and the topology of the space where
it exists. Certainly such a dependency could manifest in many ways,
typically as a function of certain topological invariants of the
space.

In the case at hand these topological invariants turn out to be the
Betti numbers of the manifold. These in turn reflect the
$\Z_2$-homology \cite{optimalGraphs} of the manifold, and so we will
now introduce very naively several concepts and illustrate them
using a well known 3-manifold, the 3-torus.

\begin{figure}
\includegraphics[width=8 cm]{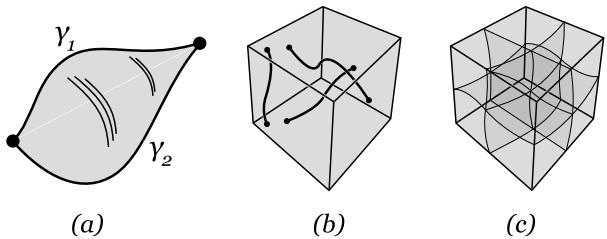}
\caption{In (a) the two curves are homologous because they form the
boundary of a deformed disc. In (b) and (c) the 3-torus is
represented as a cube in which opposite sides must be identified. In
(b) it is shown a basis for 1-cycles and in (c) a basis for
2-cycles.} \label{figura_homologia}
\end{figure}

Consider any 3-manifold $\mathcal M$. For a 1-cycle we understand
any closed non-oriented curve $\gamma$ in it, or several such
curves. In other words, it is a closed 1-manifold embedded in
$\mathcal M$. Suppose that we can embed in $\mathcal M$ a 2-manifold
in such a way that its boundary is $\gamma$. In that case $\gamma$
is called a 1-boundary and said to be homologous to zero. More
generally consider two non-oriented curves $\gamma_1$ and $\gamma_2$
with common endpoints, as in Fig.~\ref{figura_homologia} (a). We can
combine these two curves into a single 1-cycle, and then we say that
they are homologous if the 1-cycle is a 1-boundary. In other words
$\gamma_1 \sim \gamma_2$ iff $\gamma_1 + \gamma_2 \sim 0$. This kind
of equivalence can also be applied to two 1-cycles, and thus two
1-cycles are homologous iff their combination is a 1-boundary. Then
the idea is that any 1-cycle can be constructed, up to homology
equivalence, by combination of certain basic 1-cycles. The number of
1-cycles needed to form such a basis is a topological invariant, the
first Betti number $h_1$ of the manifold $\mathcal M$. For the
3-torus $h_1=3$. A possible basis in this case is the one formed by
the three 1-cycles that cross the torus in the three spatial
directions, as in fig.~\ref{figura_homologia} (b).

Similarly we can think in 2-cycles as closed 2-manifolds embedded in
$\mathcal M$. Then, when a 2-cycle is the boundary of some embedded
3-manifold it is called a 2-boundary and said to be homologous to zero.
Two 2-manifolds with common boundary can be sewed together to form a
2-cycle, and they are homologous if this 2-cycle is a 2-boundary. As
in the case of 1-cycles, there exist a basis for 2-cycles up to
homology. Again these can be exemplified in the case of a 3-torus,
see Fig.~\ref{figura_homologia}(c). The topological invariant that
gives the cardinality of such a basis is the second Betti number
$h_2$ and equals $h_1$.

Throughout the text we use sometimes a more suggestive language.
Instead of curves we will talk about strings, closed or open with
endpoints. Similarly, we will refer to embedded 2-manifolds as
membranes, either closed or with a boundary.

\begin{figure}
\includegraphics[width=8 cm]{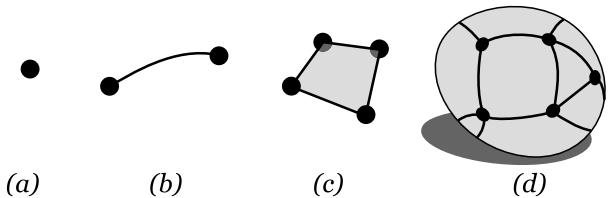}
\caption{A vertex (a), an edge (b), a face (c) and a polyhedral
solid (d).} \label{figura_cells}
\end{figure}

\subsection{System and Hamiltonian}
\label{seccionIIB}

Consider a 3-dimensional closed connected manifold $M$ that has been
constructed by gluing together polyhedral solids. These polyhedral
solids are balls whose boundary surface is a polyhedron, i.e., a
sphere divided into faces, edges and vertices, see
Fig.~\ref{figura_cells}. This gluing of polyhedral solids must
respect this structure. For brevity we will call polyhedral solids
simply as cells. Thus we have a 3-manifold divided into vertices
$V$, edges $E$, faces $F$ and cells $C$. Such a structure in a
3-manifold is called a 3-complex.

\begin{figure}
\includegraphics[width=6 cm]{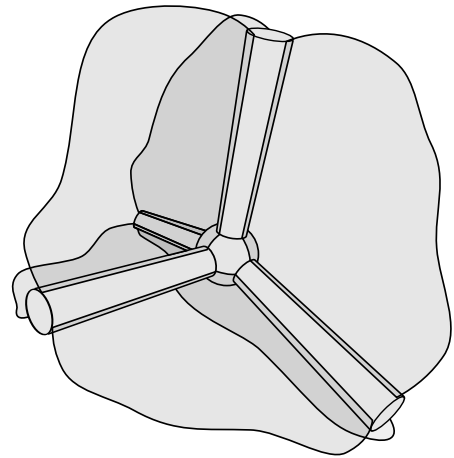}
\caption{The neighborhood of a vertex in a 3-colex. 4 edges, 6 faces
and 4 cells meet at each vertex.} \label{figura_vertice}
\end{figure}

In order to construct the topogical quantum system that we propose,
we consider a 3-complex such that

\noindent i) the neighborhood of every vertex is as the one in
Fig.~\ref{figura_vertice} and

\noindent ii) cells are four-colored, in such a way that adjacent
cells have different colors.

\noindent The colors we shall use are red, green, blue and yellow
($r$,$g$,$b$,$y$). With these assumptions we will proceed to color
edges and faces, and finally we will see that the whole structure of
the manifold is contained in the coloring of the edges.

\begin{figure}
\includegraphics[width=8 cm]{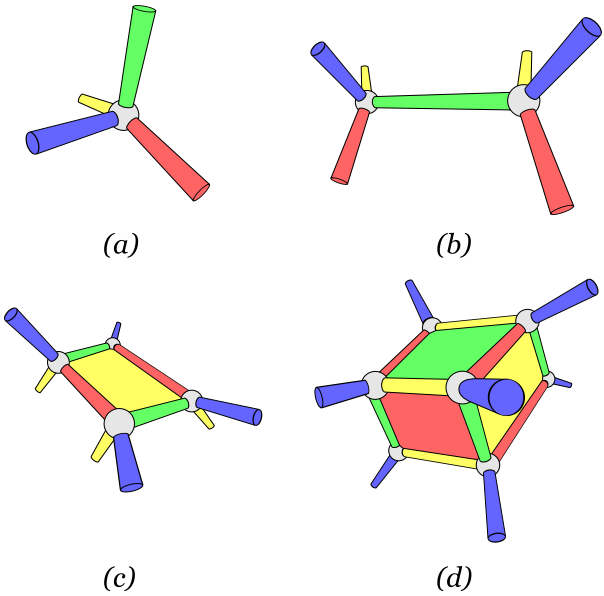}
\caption{Neighborhoods in a 3-colex of a vertex (a), a $g$-edge (b),
a $by$-face, with the yellow side visible and the blue one hidden
(c) and a $b$-cell (d). Faces are colored according to the color of
the cell at their visible side.} \label{figura_colex}
\end{figure}

With a glance at Fig.~\ref{figura_vertice} we see that the four
cells meeting at each vertex must have different colors. In the
figure we also see that each edge lyes in three cells of different
colors. Then each of the endpoints of the edge is in the corner of a
cell of a fourth color, so that we can say that it connects two
cells of the same color. We proceed to label edges with the color of
the cells they connect, see figure Fig.~\ref{figura_colex} (b). As a
result, the four edges that meet at a vertex have all different
colors, see Fig.~\ref{figura_colex} (a). Also, the edges lying on a
$r$-cell are not $r$-edges. But much more is true. Consider a
$r$-cell $c$, and any vertex $v$ in its boundary. The red edge that
ends in $v$ does not lye on the cell $c$, so that the other three
edges incident in $v$ do. But then any connected collection of $g$-,
$b$- and $y$-edges corresponds exactly to the set of edges of some
$r$-cell.

We label faces with two colors. If a face lyes between a $p$-cell
and a $q$-cell, we say that it is a $pq$-face, see
Fig.~\ref{figura_colex} (c). Then consider for example a $ry$-cell.
Since neither $r$- nor $y$-edges can lye on its boundary, this must
consist of a sequence of alternating $b$- and $g$-edges. Conversely,
any such path is the boundary of some $ry$-face. To check this,
first note that exactly one such path traverses any given $g$-edge
$e$. But $e$ must lye exactly on one $ry$-face, the one that
separates the $r$- and the $y$-cell it lyes on.

As promised, we have shown that the entire structure of the manifold
is contained in two combinatorial data: the graph and the colors of
its edges. We call the resulting structure a 3-colex, for color
complex in a 3-manifold. The simplest example of such a 3-colex with
non-trivial homology is displayed in Fig.~\ref{figura_p3}. It
corresponds to the projective space $P^3$. In appendix
\ref{apendice_inflado} we will give a procedure to construct a colex
of arbitrary dimension $D$, or $D$-colex, starting with an arbitrary
complex in a $D$-manifold.

\begin{figure}
\includegraphics[width=6 cm]{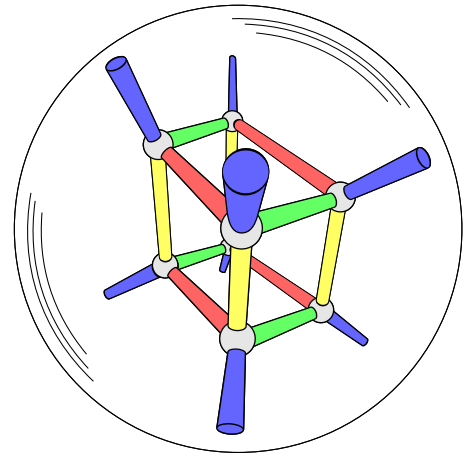}
\caption{The projective space $P^3$ can be obtained starting with a
solid sphere and identifying opposite points in its surface. Here we
use such a representation to show a 3-colex in
$P^3$.}\label{figura_p3}
\end{figure}

We now associate a physical system to the 3-colex. To this end, we
place at each vertex (site) a spin-$\frac 1 2$ system. To each cell
$c$, we attach the cell operator
\begin{equation}\label{operador_celda}
B_c^X := \bigotimes_{i\in I_c} X_i,
\end{equation}
where $X_i$ is the Pauli $\sigma_1$ matrix acting on site $i$ and
$I_c$ is the set of sites lying on the cell. Similarly, to each face
$f$ we attach the face operator
\begin{equation}\label{operador_cara}
B_f^Z := \bigotimes_{i\in I_f} Z_i,
\end{equation}
where $Z_i$ is the Pauli $\sigma_3$ matrix acting in site $i$ and
$I_f$ is the set of sites lying on the face. We have
\begin{equation}\label{commutacion}
\paratodo c\in C, f\in F, \qquad [B_c^X,B_f^Z]=0.
\end{equation}
To show this, consider any cell $c$ and face $f$. The edges of $c$
come in three colors and the edges of $f$ in two. Thus they have a
least a common color, say $q$. Given any shared vertex, we consider
its $q$-edge $e$. But $e$ lyes both on $c$ and $f$, and thus its
other endpoint is also a shared vertex. Therefore $c$ and $f$ share
an even number of vertices and $[B_c^X, B_f^Z]=0$.

The Hamiltonian that we propose is constructed by combining cell and
face operators:
\begin{equation}\label{Hamiltoniano}
H= - \sum_{c\in C} B_c^X - \sum_{f\in F} B_f^Z
\end{equation}
Observe that color plays no role in the Hamiltonian, rather, it is
just a tool we introduce to analyze it. In appendix
\ref{apendice_contaje} we calculate the degeneracy of the ground
state. It is $2^k$ with
\begin{equation}\label{degeneracion}
k=3h_1,
\end{equation}
and therefore depends only upon the manifold, which is a signature
of topological quantum order.

 The ground states $\ket \psi$ are
characterized by the conditions
\begin{alignat}2
 \paratodo c\in C\quad\quad &B_c^X \ket
\psi &= \ket \psi, \label{condiciones_cell}\\
\paratodo f \in F\quad\quad &B_f^Z \ket
\psi &= \ket \psi, \label{condiciones_face}
\end{alignat}
for cell and face operators. Those eigenstates $\ket {\psi'}$ for
which any of the conditions is violated is an excited state. There
are two kinds of excitations. If $B_c^X \ket {\psi'} = - \ket
{\psi'}$ we say that there is an excitation at cell $c$. Similarly,
if $B_f^Z \ket {\psi'} = - \ket {\psi'}$ then the face $f$ is
excited. Below we will show that cell excitations are related to
quasiparticles and face excitations to certain flux. For know we are
just interested in noting that excitations have a local nature and
thus the Hamiltonian \eqref{Hamiltoniano} is gapped. Then since the
ground state degeneracy depends upon the topology, we have
topological quantum order.

\subsection{Strings and Membranes}
\label{seccionIIC}

From this point on we shall pursue a better understanding both the
ground state degeneracy and the excitations by means of the
introduction of string and membrane operators. In this direction, an
essential notion will be that of a \emph{shrunk complex}, both of
the first and second kind. The motivation after the construction of
these complexes from the colexes is that only at the shrunk complex
level it is possible to visualize neatly the strings and membranes
that populate the model. These new shrunk complexes are not colexes,
but its cells are associated to cells in the colex, and thus have
color labels.

\subsubsection{Shrunk complex of the first kind}
\begin{figure}
\includegraphics[width=8 cm]{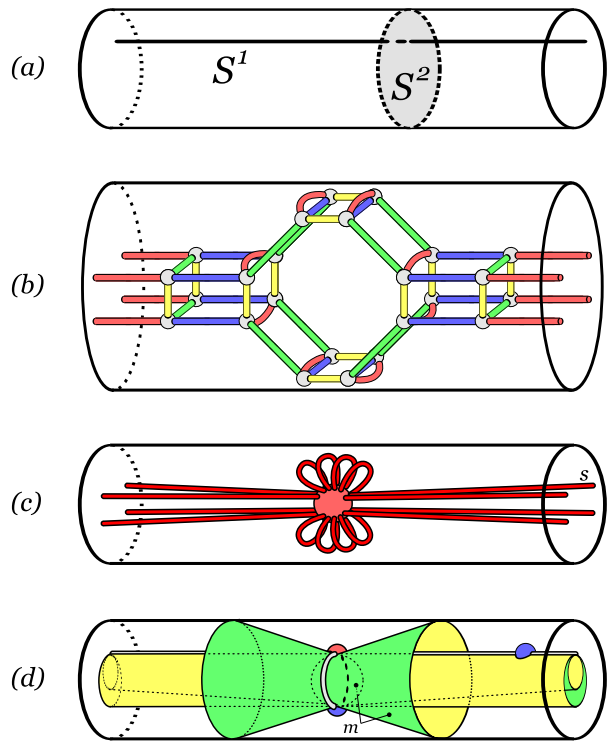}
\caption{(a) A representation of the space $S^2\times S^1$. Each
section of the solid tube is a sphere, and both ends of the tube are
identified. (b) A 3-colex in $S^2\times S^1$. It consists of 24
vertices, 12 edges of each color, 4 $br$-faces, 8 $by$-faces, 6 
$rg$-faces, 4 $ry$-faces, 4 $gy$-faces, 2 $b$-cells, 1 $r$-cell, 3
$g$-cells and 2 $y$-cells. (c) The $r$-shrunk complex of the
previous colex. The vertex corresponds to a $r$-cell, and edges to
$r$-edges. An example of closed string is the edge marked with a
$s$. It has nontrivial homology. (d) The $gy$-shrunk complex of the
previous colex. Vertices correspond to $b$- and $r$-cells, edges to
$rb$-faces, face to $gy$-faces and cells to $g$- and $y$-cells. An
example of a closed membrane is the combination of the faces marked
with a $m$. This membrane has nontrivial
homology.}\label{figura_shrunk}
\end{figure}

\begin{figure}
\includegraphics[width=7 cm]{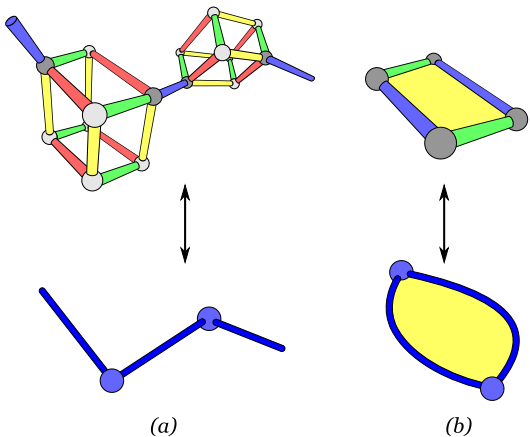}
\caption{In this figure the top represents part of a colex and the
bottom the corresponding portion of the $b$-shrunk complex. Vertexes
in the $b$-shrunk complex come from $b$-cells in the colex, edges
from $b$-edges, faces from $ry$-, $rg$- and $gy$-faces and cells
from $r$-, $y$- and $g$-cells. (a) A $b$-string. In the colex it is
a collection of $b$-edges linking $b$-cells. In the $b$-shrunk
complex the path of edges can be clearly seen. (b) A $ry$-face
corresponds to a face in the $b$-shrunk complex, and thus its
boundary can be viewed as a $b$-string.}\label{figura_string}
\end{figure}

It is associated to a color, and it allows to visualize strings of
that particular color. Consider for example the $b$-shrunk complex.
The idea is that we want to keep only $b$-edges, whereas $g$-,$r$-
and $y$-edges get shrunk and disappear. To this end, we start
placing a vertex at each $b$-cell and connecting them through edges,
which are in one to one correspondence with $b$-edges. Then we have
to place the faces of the new complex, and they correspond to $rg$-,
$ry$- and $gy$-faces. In particular, consider a $rg$-face. It has
$b$- and $g$-edges, but after $g$-edges are shrunk only $b$-edges
remain. Finally we need cells. They come from $g$-, $r$- and
$y$-cells. In particular, consider a $g$-cell. It has $r$-, $y$- and
$b$-edges, but only $b$-edges are retained. Similarly, it has $gb$-,
$gr$- and $gy$-faces, but we keep only $gb$-faces. See
Fig.~\ref{figura_shrunk} (c) for an example an also
Fig.~\ref{figura_string}.

Now consider any path, closed or not, in the $b$-shrunk complex. We
call such a path a $b$-string. Recall that each edge of a shrunk
complex corresponds to a $b$-edge in the 3-colex. Thus at the colex
level a $b$-string is a collection of $b$-edges that connect
$b$-cells, see Fig.~\ref{figura_string} (a). Each $b$-edge contains
two vertices. Then to each $b$-string $s$ we can associate an
operator $ B_s^Z = \bigotimes_{i\in I_s} Z_i,$ where $I_s$ is the
set of vertices lying in the string.

As shown in Fig.~\ref{figura_string} (b), the operator $B_f^Z$ of a
$yr$-face corresponds to a closed $b$-string $s$. This string is the
boundary of the corresponding face in the $b$-shrunk complex. As an
operator, $B_s^Z$ clearly commutes with the Hamiltonian and acts
trivially on the ground state \eqref{condiciones_cell}.

In fact, any closed string gives rise to a string operator that
commutes with the Hamiltonian\eqref{Hamiltoniano}. If the string is
homologous to zero the corresponding string operator acts trivially
on the ground state. In order to understand this, consider a closed
red string homologous to zero. It must be a combination of
boundaries of faces. Then the string operator is the product of the
operators of these faces. Similarly, the action of two string
operators derived from homologous strings of the same color is
identical on the ground state. Therefore it makes sense to label the
string operators as $S_\mu^{p}$, where $p$ is a color and $\mu$ is a
label denoting the homology of the string.

\subsubsection {Shrunk complex of the second kind} 
\begin{figure}
\includegraphics[width=7 cm]{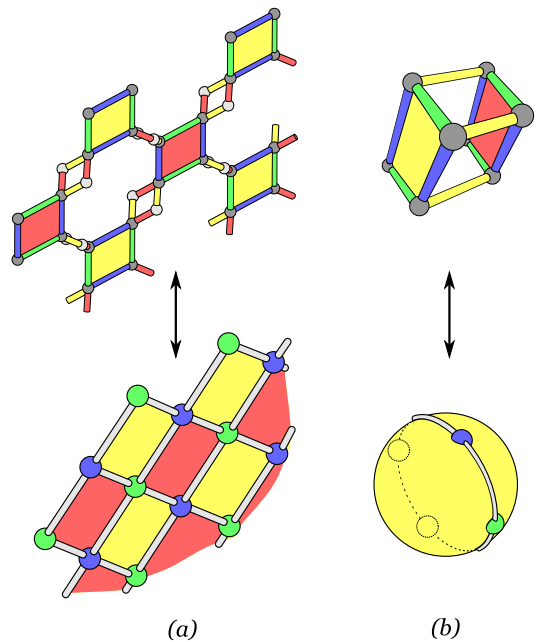}
\caption{In this figure the top represents part of a colex and the
bottom the corresponding portion of the $ry$-shrunk complex.
Vertexes in the $ry$-shrunk complex come from $g$- and $b$-cells in
the colex, edges from $gb$-faces, faces from $ry$-faces and cells
from $r$- and $y$-cells.(a) A $ry$-membrane. In the colex it is a
collection of $ry$-faces linked by $gb$-faces. In the $ry$-shrunk
complex the brane can be clearly seen (b) A $r$-cell corresponds to
a cell in the $ry$-shrunk complex, and thus its boundary can be
viewed as a $ry$-membrane. }\label{figura_membrane}
\end{figure}

It is associated to
two colors, and it allows the visualization of certain membranes, as
we explain now. Let us consider for example the $ry$-shrunk complex.
The idea is that we want to keep only $ry$-faces, whereas the rest
of faces get shrunk and disappear. This time vertices correspond to
$b$- and $g$-cells. Edges come from $bg$-faces. A $bg$-face lyes
between a $g$- and a $b$-cell, and the corresponding edge will
connect the vertices coming from these cells. We have already
mentioned that the faces of the $ry$-shrunk complex come from
$ry$-faces in the colex, but we have to explain how they are
attached. Observe that each
 $ry$-face has certain amount of adjacent $gb$-faces. Here for
 adjacent objects
we will only mean that their intersection is not empty. In
particular there is a $gb$-face at each of the vertices of the
$ry$-face. Then the face in the complex has in its perimeter the
edges coming from its adjacent $gb$-faces. Finally we have to
consider cells, which come from $r$- and $y$-cells and only keep
their $ry$ faces. So in the boundary of a cell coming from an
$r$-cell we see vertices from adjacent $b$- and $g$-cells, edges
from adjacent $bg$-faces and faces from $ry$-faces in the boundary
of the $r$-cell. See Fig.~\ref{figura_shrunk} (d) and
Fig.~\ref{figura_membrane}.

Now consider any membrane, that is, a connected collection of faces,
closed or with a boundary, in the $ry$-shrunk complex. We call such
a membrane $m$ a $ry$-membrane, see Fig.~\ref{figura_shrunk} (d) and
Fig.~\ref{figura_membrane} (a). We can associate an operator $B_m^X$
to it. It is the product of the $B_f^X$ operators of the
corresponding $ry$-faces in the colex.

As shown in Fig.~\ref{figura_membrane} (b), the operator $B_c^X$ of
a $r$-cell $c$ correspond to a closed $ry$-membrane $m$. This
membrane is the boundary of the corresponding cell in the
$ry$-shrunk complex. As an operator, $B_m^X$  clearly commutes with
the Hamiltonian and acts trivially on the ground state
\eqref{condiciones_face}.

In complete analogy with strings, any closed membrane gives rise to
a membrane operator that commutes with the Hamiltonian. If the
membrane is homologous to zero then the corresponding membrane
operator acts trivially on the ground state. Similarly, the action
of two string operators derived from homologous membranes of the
same color is identical on the ground state, and we label membrane
operators as $M_\mu^{pq}$, where $p$ and $q$ are colors and $\mu$ is
a label denoting the homology of the membrane.

\subsubsection{Commutation rules}
\begin{figure}
\includegraphics[width=7 cm]{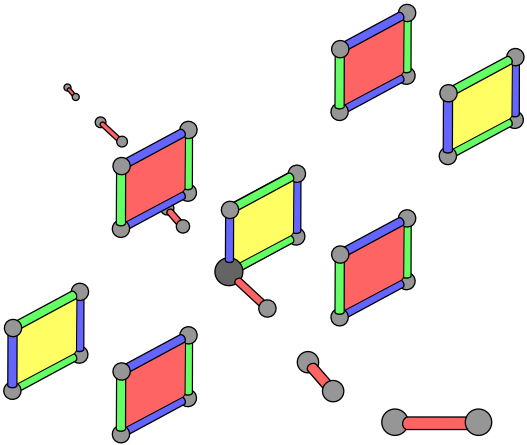}
\caption{When a $r$-string $s$ crosses a $ry$-membrane $m$, they
meet at a vertex. In terms of string and membrane operators, this
means that $B_m^X$ and $B_s^Z$ act in a common
site.}\label{figura_interseccion}
\end{figure}

We will now consider the commutation rules between string and
membrane operators. We first consider the case of a membrane and a
string with no common color in their labels. As displayed in figure
Fig.~\ref{figura_membrane}(a), a $rg$-membrane is made up of $g$-
and $b$-edges. Then for the same argument of \eqref{commutacion} we
have
\begin{equation}\label{}
\paratodo \mu, \nu\quad\quad[M_\mu^{rg},S_\nu^b]=0,
\end{equation}
and analogously for any combination of three different colors. More
interesting is the case in which there is a shared color. As
displayed in Fig.~\ref{figura_interseccion}, at each place where a
$p$-string crosses a $pq$-membrane they have a site in common. Thus,
if the labels $\mu$ and $\nu$ are such that a $\nu$ string crosses a
$\mu$ membrane an odd number of times, we have
\begin{equation}\label{}
\{M_\mu^{pq},S_\nu^p\}=0.
\end{equation}
In other case, that is, if they cross an even number of times,
the operators commute.

\subsection{Ground state}
\label{seccionIID}

 Above we have discussed how the action of string
or membrane operators on the ground state depends only upon their
homology. It is in this sense that homologous strings or membranes
give rise to equivalent operators. This equivalence however can be
extended to take color into account, and we say that two membrane or
string operators are equivalent if they are equal up to combinations
with cell and face operators. Then, as we prove for general $D$ in
appendix \ref{apendice_combinacion}, we have the following interplay
between homology and color:
\begin{align}
S_\mu^{r}S_\mu^{g}S_\mu^{b}S_\mu^{y}&\sim 1,\label{combina_cuerdas} \\
M_\mu^{pq}M_\mu^{qo}M_\mu^{op}&\sim 1,\label{combina_membranas}
\end{align}
where $o$, $p$ and $q$ are distinct colors.

\begin{figure}
\includegraphics[width=7 cm]{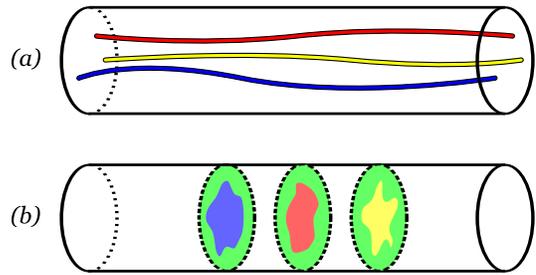}
\caption{Here we represent $S^2\times S^1$ as in
Fig.~\ref{figura_shrunk}. In (a) it is shown a basis for nontrivial
closed strings. The other possible such string is green, but it is a
combination of these ones \eqref{combina_cuerdas}. (b) A membrane
basis in $S^2\times S^1$. We have chosen a $gb$, a $gr$ and a
$gy$-membrane. There are three other nontrivial membranes, in
particular a $br$-, a $by$- and a $yr$-membrane, but they are
combinations of these ones
\eqref{combina_membranas}.}\label{figura_base_s2xs1}
\end{figure}

If we take all the $r$-, $g$- and $b$-strings for a given homology
basis of 1-cycles, we obtain a complete set of compatible
observables for the ground state subspace: any other string operator
is equivalent to a combination of these strings, and no membrane
operator that acts nontrivially in the ground state can commute with
all of them. This is in fact why the number 3 appears in
\eqref{degeneracion}. As an example, a string basis in $S^2\times
S^1$ is displayed in Fig.~\ref{figura_base_s2xs1} (a).

Similarly, if we take all the $ry$-, $gy$- and $by$-membranes for a
given homology basis of 2-cycles, we obtain a complete set of
compatible observables for the ground state subspace: any other
membrane operator is equivalent to a combination of these membranes,
and no string operator that acts nontrivially in the ground state
can commute with all of them. A membrane basis in $S^2\times S^1$ is
displayed in Fig.~\ref{figura_base_s2xs1} (b).

\begin{figure}
\includegraphics[width=5 cm]{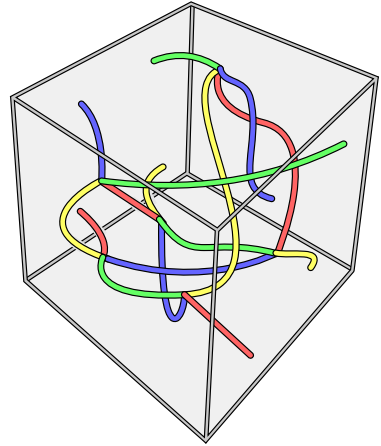}
\caption{The ground state of the system is a string-net condensate.
This picture represents in a 3-torus a typical element of the
sumation \eqref{string-net}.}\label{figura_string-net}
\end{figure}

Observe that only those string operators that have nontrivial
homology, that is, which act in a global manner in the system, are
capable of acting nontrivially in the ground state whereas living it
invariant. This is the signature of a string condensate, as
introduced in \cite{hammazanardiwen05}. Then it would be tempting to let
$S_b$ be the set of all boundary strings and try to write a ground
state as
\begin{equation}\label{}
\sum_{s\in S_b} B_s^Z {\ket \rightarrow} ^{\otimes |V|},
\end{equation}
where ${\ket \rightarrow} ^{\otimes |V|}$ is the state with all
spins pointing in the positive $x$ direction. However, this fails.
In fact, what we have is a \emph{string-net condensate}
\cite{levinwen05} because, as indicated by
\eqref{combina_cuerdas}, we can have branching points in which one
string of each color meet. This means that the ground state is a
superposition of all possible nets of strings, as depicted in
Fig.~\ref{figura_string-net}. The correct way to write an example of
a ground state is
\begin{equation}\label{string-net}
\sum_{f\in F} (1+B_f^Z) {\ket \rightarrow} ^{\otimes |V|} =:
\sum_{\text{string-nets}} B_s^Z {\ket \rightarrow} ^{\otimes |V|}.
\end{equation}

We can state all of the above also in the case of membranes, and
thus we should speak of a \emph{membrane-net condensate}, and interestingly enough,
in other topological orders in $D=3$ based on toric codes do not exhibit
a condensation of membrane-nets \cite{hammazanardiwen05}.
It is a membrane condensate because only
membranes with nontrivial homology can act nontrivially in the
ground state. And it is a net because, for example, as indicated by
\eqref{combina_membranas} a $gr$-, a $gb$- and a $br$-membrane can
combine along a curve. Then if we let ${\ket \uparrow} ^{\otimes
|V|}$ denote the state with all spins up, the following is an
example of a ground state:
\begin{equation}\label{}
\sum_{c\in C} (1+B_c^X) {\ket \uparrow} ^{\otimes |V|}=:
\sum_{\text{membrane-nets}} B_m^X {\ket \uparrow} ^{\otimes |V|}.
\end{equation}

\subsection{Excitations}
\label{seccionIIE}

\begin{figure}
\includegraphics[width=5 cm]{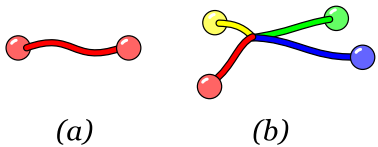}
\caption {There are two ways in which quasiparticles can be created
locally. We can either create them by pairs of the same color forming a string (a) or
in groups, one of each color forming a string-net (b).}\label{figura_particulas}
\end{figure}

We now focus on excitations from the point of view of string and
membrane operators. We can have two kinds of excitations, depending
on wether a cell or face condition is violated. We start by
considering excitations in $r$-cells, for example. Let $\ket \psi$
be a ground state and let $S_{ij}^r$ be an open string operator
connecting the cells $i$ and $j$. The state $S_{ij}^r\ket \psi$ is
an excited state. The excitations live precisely at cells $i$ and
$j$, and we call them quasiparticles with $r$-charge. Why should be
color be considered a charge? We have the following 3 constraints:
\begin{equation}\label{condicionesX}
\prod_{c\in C_r} B_c^{X}= \prod_{c\in C_g} B_c^{X} =\prod_{c\in C_b}
B_c^{X} = \prod_{c\in C_y} B_c^{X},
\end{equation}
where $C_p$ is the set of $p$-cells. They imply that the number of
quasiparticles of each color must agree in their evenness or
oddness. Therefore, if we want to create quasiparticles of a single
color from the vacuum we must create them in pairs, and so such a
creation can be performed with an open string operator.
Alternatively, 4 quasiparticles, one of each color, can also be
created locally, see Fig.~\ref{figura_particulas} (b). For example,
let $\ket \psi$ be a ground state and $i$ any site. Then the state
$Z_i\ket \psi$ is an state with four quasiparticle excitations, one
at each of the 3-cells that meet at site $i$. Observe that
\eqref{condicionesX} is in agreement with
\eqref{combina_cuerdas}.

\begin{figure}
\includegraphics[width=6 cm]{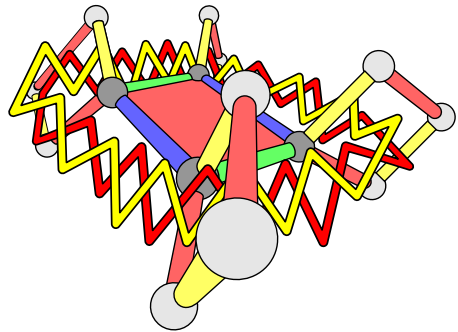}
\caption{The flux excitation created with the membrane operator
$B_m^X$ of a $ry$-membrane made up of a single
$ry$-face.}\label{figura_flujo_membrana}
\end{figure}

\begin{figure}
\includegraphics[width=8 cm]{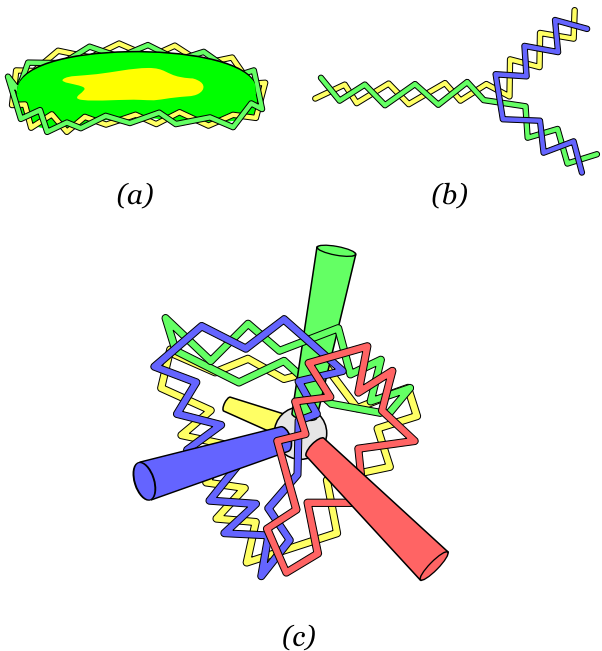}
\caption {(a) The border of a $gy$-membrane is a $gy$-flux. (b) A
$gy$-flux can split in a $gb$-flux and a $by$-flux when it goes
across a $r$-cell \eqref{condicionesZ}. (c) The microfluxes at a
given site, as explained in the text.} \label{figura_flujos}
\end{figure}

Now let $\ket \psi$ be a ground state and let $M_{b}^{gy}$ be a
membrane operator which has a boundary $\partial b$. Recall that
$\partial b$ is a set of edges in the $gy$-shrunk complex that
corresponds to a set of $rb$-faces at the colex level. The state
$M_b^{gy}\ket \psi$ is an excited state with excitations placed at
the faces in $\partial b$. The excited segments, as viewed in the
$gy$-shrunk complex, form a closed path. This motivates the idea of
a $gy$-flux in the boundary of the membrane, as illustrated in
Fig.\ref{figura_flujo_membrana} for a membrane with a single face.
But we have to check that this flux makes sense. Not only it must be
conserved at any vertex in the $gy$-shrunk graph, but also the
existence of fluxes of other colors must be considered. So take for
example a $r$-cell $c$. We have 2 constraints for the faces of $c$,
analogous to those in \eqref{condicionesX} but in the subcolex that
forms the boundary of $c$:
\begin{equation}\label{condicionesZ}
\prod_{f\in F_{rb}^c} B_f^{Z} = \prod_{f\in F_{rg}^c} B_f^{Z} =
\prod_{f\in F_{rb}^c} B_f^{Z},
\end{equation}
where $F_{pq}^c$ is the set of $pq$-faces of the cell $c$. These
constraints guarantee that $gy$-flux is preserved at the
corresponding vertex in the $gy$-shrunk complex. Additionally,
\eqref{condicionesZ} imply that a $gy$-flux can split in a $gb$-flux
plus a $yb$-flux, see Fig.~\ref{figura_flujos} (b). This is of
course in agreement with \eqref{combina_membranas}.

Fluxes can be analyzed from a different point of view. Let $\ket
\psi$ be a ground state and $i$ any site. Then the state $X_i\ket
\psi$ is an excited state. We can visualize it as small $p$-fluxes
winding around the $p$-edges incident at $i$, as shown in
Fig.~\ref{figura_flujos} (c). Observe that the idea of a $pq$-flux
as something composed of a $p$-flux and a $q$-flux is also suggested
by the flux splitting \eqref{condicionesZ}. Any flux configuration
is a combination of these microfluxes at sites. In particular, the
total flux through any closed surface must be null, and thus we
cannot have, for example, an isolated $rg$-flux in a loop which is
not homologous to zero.

\subsection{Winding quasiparticles around fluxes}
\label{seccionIIF}

\begin{figure}
\includegraphics[width=5 cm]{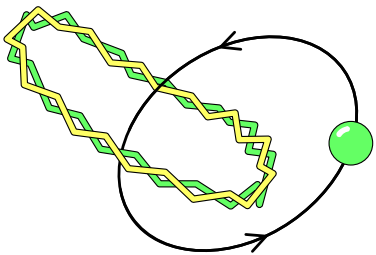}
\caption {When a $g$-charge winds around a loop of $gy$-flux the
system gets a global -1 phase. This is because the membrane operator
giving rise to the flux and the string operator associated to the
winding anticommute.} \label{figura_winding}
\end{figure}

In the theory of topological order in 2D it is known that
quasiparticles show special statistics \cite{wilczek82}, \cite{leinaasmyrheim77}:
when a charge is
carried around another one, sometimes the system gets a global
phase, a behavior which bosons and fermions do not show. Which is
the analogous situation in 3D? We can carry a charged particle along
a closed path which winds around a loop of flux, as in
Fig.~\ref{figura_winding}. If the system gets a global phase, then
it makes sense to introduce the notion of branyons as the higher
dimensional generalization of the usual anyons. Thus in the system
at hand we have 0-branyons (quasiparticles) and 1-branyons (fluxes).
Higher dimensional branyons will appear when we consider systems
with $D\geq 4$.

In order to see the effect of winding a color charge around a color
flux, we have to consider the closed string operator associated to
the charge path and the membrane giving rise to the flux loop. If a
$p$-charge winds once around a $pq$-flux, the system will get a
global $-1$ phase because $\{M^{pq},S^p\}=0$. Observe that
this reinforces the idea of a $pq$-flux as a composition of a
$p$-flux and a $q$-flux. Other color combinations, i.e., those in
which string and membrane do not share a color, give no phase.

\section{$D$-colexes}
\label{seccionIII}

In order to generalize the 3-dimensional model to higher dimension
$D$ we have first to construct the underlying structure. That is, we
want to define color complexes of arbitrary dimension. This section
is devoted to the definition and basic properties of $D$-colexes.

\subsection{Definitions}
\label{seccionIIIA}

 First we define color graphs or \emph{c-graphs}. A $v$-valent c-graph
 is a graph $ \Gamma$ satisfying that

\noindent \emph{(i)}  $v$ edges meet at every vertex,

\noindent \emph{(ii)} edges are not loops and

\noindent \emph{(iii)} edges are $v$-colored.

\noindent For being $v$-colored we mean that labels from a color set
$Q=\sset{q_1,\dots,q_v}$ have been assigned to edges in such a way
that two edges meeting at a vertex have different colors. This is a
generalization of what we already saw in the $D=3$ case, as in
Fig.~\ref{figura_colex}. A c-graph $\Gamma'$ with color set $Q'$ is
a c-subgraph of $\Gamma$ if $\Gamma'\subset\Gamma$, $Q'\subset Q$
and the colorings coincide in common edges.

Now we introduce \emph{complexes}. One can give to a $D$-manifold a
combinatorial structure by means of what is called a $D$-complex.
The idea is to divide the manifold in a hierarchy of objects of
increasing dimension: points, edges, faces, solid spheres, etc.
These objects are called $n$-cells, $n=0,\dots, D$. 0-cells are
points, 1-cells are edges, and so on. The boundary of a $n$-cell is
a $n$-sphere and is made up of cells of dimension $n'<n$. So what we
have is a $D$-manifold constructed by gluing together the higher
dimensional analogs of the polihedral solids that we considered in
$D=3$, recall Fig.~\ref{figura_cells}.

A $D$-colex is a complex in a $D$-manifold which has $(D+1)$-colored
edges in such a way that

\noindent \emph{(i)} its underlying lattice or graph is a (D+1)-valent c-graph,

\noindent \emph{(ii)} the subgraph that lyes on the boundary of any
$n$-cell for $n=2,\dots,D$ is a $n$-valent c-subgraph and

\noindent \emph{(iii)} any connected c-subgraph with valence $v=2,\dots,D$
lyes on the boundary of one unique $v$-cell.

\noindent Therefore the point is that the c-graph \emph{completely}
determines the cell structure and thus the whole topology of the
manifold.

Some c-graphs yield a colex, but not all of them. We define
recursively this partially defined mapping from the space of
$(D+1)$-valent c-graphs to the space of closed $D$-manifolds. First,
any 2-valent c-graph is a collection of loops, so as a topological
graph it naturally yields a 1-manifold, namely a collection of
1-spheres. Then consider any 3-valent c-graph. We construct a
2-complex starting with the corresponding topological graph or
1-complex. The idea is first to list all 2-valent c-subgraphs, which
are embeddings of $S^1$ in the 1-complex. Then for each of these
subgraphs we attach a 2-cell, gluing its boundary to $S^1$.
The resulting space is certainly a 2-manifold. It is enough to check
a neighborhood of any vertex, but the one to one correspondence
between cells and connected c-subgraphs makes this straightforward.
Then we consider a 4-valent c-graph. If not all of its 3-valent
c-subgraphs yield $S^2$, we discard it. Otherwise we first proceed
to attach 2-cells as we did for the 3-valent graph. Then we list all
3-valent c-subgraphs, which by now correspond to embeddings of $S^2$
in a 2-complex. At each of these spheres we glue the surface of a
solid sphere. The process can be continued in the obvious way and
thus in general a $(D+1)$-valent c-graph yields a $D$-colex iff all
its $D$-valent c-subgraphs yield $S^D$.

\subsection{Examples}
\label{seccionIIIB}

\begin{figure}
\includegraphics[width=5 cm]{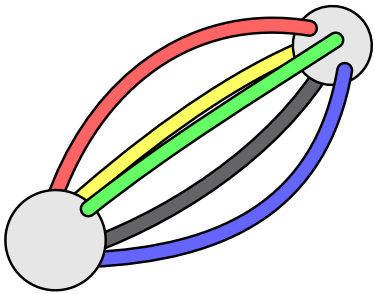}
\caption {A c-graph that yields a 4-sphere. It is the simplest
possible 4-colex, with only two vertices.} \label{figura_esfera}
\end{figure}

As a first example of colex, consider the c-graph composed of only
two vertices, for any valence $v=D+1\geq 2$. An example can be
found in Fig.~\ref{figura_esfera}. This family of c-graphs yields
the spheres $S^D$. This can be visualized viewing $S^D$ as $\R^D$
plus the point at infinity. We can place one vertex at the origin
and the other at infinity. Then edges are straight lines that
leave the origin in different directions.

The projective space $P^D$ can also be described easily with a
colex, though less economically in terms of vertices. Recall that
$P^D$ can be constructed by identifying opposite points of the
boundary of a $D$-dimensional ball. The idea is to consider a
$D$-cube and construct a $D$-valent $c$-graph with its vertices and
edges, coloring parallel edges with the same color. Then we add
$2^{n-1}$ extra edges to connect opposite vertices, and give them a
new color. The resulting c-graph yields $P^D$. See
Fig.~\ref{figura_p3} for an example in the case $D=3$.

In appendix \ref{apendice_inflado} we give a procedure to construct
colexes from arbitrary complexes. This guarantees that we can
construct our topologically ordered physical system in any closed
manifold with $D\geq 2$.

\subsection{R-shrunk complex}
\label{seccionIIIC}

This section is devoted to the construction of several new complexes
from a given colex. These constructions will be essential for the
understanding of the physical models to be built. In particular, as
we learnt in the $D=3$ case, only at the shrunk complex level will
it be possible to visualize neatly the branes that populate the
system.
Shrunk complexes also provide us with several relations
among the cardinalities of the sets $C_n$ of $n$-cells, which in
turn will be essential to calculate the degeneracy of the ground
state. These relations are based on the Euler characteristic of a
manifold, a topological invariant defined in a $D$-complex as:
\begin{equation}\label{Euler}
\chi := \sum_{n=0}^D (-1)^n |C_n|
\end{equation}

Before starting with the construction, it is useful to introduce the
notion of the Poincar\'e dual of a complex $\mathcal C$ in a
$D$-manifold. The dual complex $\mathcal C^\ast$ is obtained by
transforming the $n$-cells of $\mathcal C$ in $(D-n)$-cells, and
inverting the relation being-a-boundary-of. This means that if
certain $(n-1)$-cell $c'$ is in the boundary of the $n$-cell $c$ in
$\mathcal C$, then $c^\ast$ is in the boundary of $c'^\ast$ in
$\mathcal C^\ast$.

We say that a cell is a $R$-cell if its c-graph has as color set
$R$. Note that this notation is different from the one we used in
$D=3$, but it is more suitable for high $D$. What before was a
$gy$-membrane now will be a $\sset{r,b}$-brane, or more simply a
$br$-brane, and so on.

Consider a $D$-colex $\mathcal C$ with color set $Q$. We want to
construct its $R$-shrunk complex $\mathcal C_R$, where $R$ is a
nonempty proper subset of $Q$, $\emptyset \subsetneq R\subsetneq Q$.
What we seek is a new complex in which only $R$-cells remain whereas
the rest of $|R|$-cells disappear. This construction is accomplished
by a partial Poincar\'e dualization of cells. We already saw
examples of this construction in $D=3$. Due to the different
notation, a $gy$-shrunk complex there will be a $rb$-shrunk complex
here.

\begin{figure}
\includegraphics[width=6 cm]{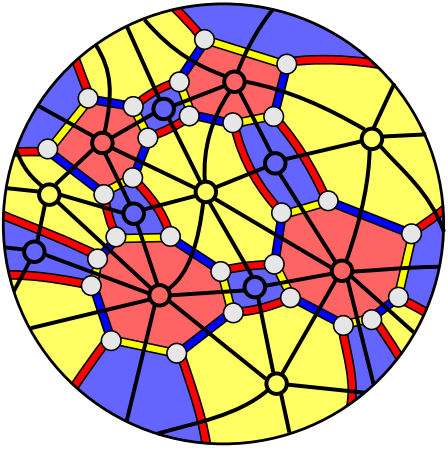}
\caption {A $bry$-cell belonging to some $D$-colex with $D\geq 3$.
Superimposed we show in black thick line the structure of its dual boundary, which plays
an important role when constructing the $bry$-shrunk complex.}
\label{figura_dualizacion}
\end{figure}

\begin{figure}
\includegraphics[width=8 cm]{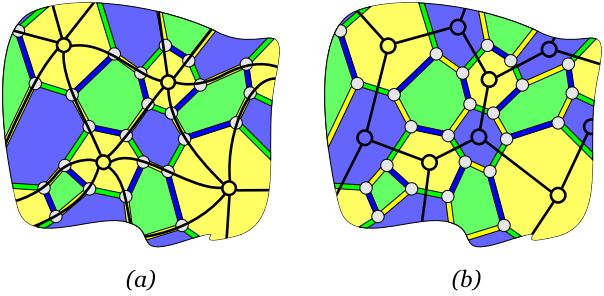}
\caption {This figure shows the two possible kinds of shrunk complex
in a 2-colex. The shrunk complexes appear superimposed in black thick line to the
original colex. In (a) it is shown the $y$-shrunk complex, and in
(b) the $by$-shrunk complex.} \label{figura_shrunk2d}
\end{figure}

The $R$-shrunk complex has two main sets of cells. The first one
corresponds to the cells in the set
\begin{equation}\label{}
S_1:=\bigcup_{R \subseteq S \subsetneq Q} C_S,
\end{equation}
where  $C_S$ is the set of $S$-cells. Cells in $S_1$ keep their
dimension and the relation being-the-boundary-of among them. The
second cell set is
\begin{equation}\label{}
S_2:=\bigcup_{\bar R \subseteq S \subsetneq Q} C_S,
\end{equation}
where $\bar R$ is the complement of $R$ in $Q$
\begin{equation}\label{R_barra}
\bar R:=Q-R.
\end{equation}
Cells in $S_2$ get dualized. This means that a $n$-cell in the colex
will be a $(D-n)$-cell in the $R$-reduced complex. The relation
being-the-boundary-of is inverted among the cells in $S_2$. So $S_2$
provides us with cells of dimensions 0,\dots, $|R|-1$ and $S_1$ with
cells of dimensions $|R|,\dots, D$. Up to dimension $|R|-1$ the
construction is clear, but we have to explain how to attach the
cells in $S_1$. To this end, we observe that the intersection of a
$n$-cell in $S_1$ and a $R$-cell is either empty or a cell of
dimension $n' = n-|\bar R|$. The $n$-cell gives rise to a cell of
dimension $D-n=|R|-1-n'$. Thus, the partial dualization is in fact a
complete dualization as seen on the boundary of any $R$-cell, and
the attachment of each $R$-cell is then naturally described by this
dualization process, as shown in Fig.~\ref{figura_dualizacion}. For
the cells coming from $S$-cells with $R\subsetneq S$ the attachment
can be described recursively. The boundary of these cells is a
$(|S|-1)$-colex, so we can obtain its $R$-shrunk complex and use it
as the new boundary for the cell. In fact, what we are doing is a
projection of the shrunking process in the boundary of the cell.
Fig.~\ref{figura_shrunk2d} displays examples of shrunk complexes for
$D=2$.

 The Euler characteristic for a $R$-shrunk complex is
 \begin{equation}\label{}
 \chi =  \sum_{R \subseteq S \subsetneq Q } (-1)^{|S|} |C_S|
 + \sum_{\bar R \subseteq S \subsetneq Q } (-1)^{D-|S|} |C_S|.
 \end{equation}
If we sum up all such equations for all different color combinations
but fixed cardinality $|R|=r$ we get
\begin{multline}\label{Euler_plus}
\binom {D+1} {r} \chi =  \sum_{n=r}^{D} (-1)^n \binom n {r} |C_n| + \\
+ \sum_{n=0}^{r-1} (-1)^n \binom {D-n} {D-r+1} |C_{D-n}|.
\end{multline}
The case $r=0$ is also included since it reduces to the definition
of $\chi$. The rhs are equal in the cases $r=s$ and $r=D-s+1$ except
for a sign, so that we get
\begin{equation}\label{}
\chi = (-1)^{D}\chi.
\end{equation}
Of course, this is the well known fact that $\chi$ vanishes in
manifolds of odd dimension. In these cases in which $D=2k+1$,
equation \eqref{Euler_plus} for $r=k+1$ vanishes identically. So in
general we have $\lceil D/2 \rceil$ independent relations. They tell
us that the cardinalities $|C_0|,\dots,|C_{\lfloor D/2\rfloor}|$
depend of the cardinalities $|C_{\lfloor
D/2+1\rfloor}|,\dots,|C_D|$, which shows quantitatively the fact
that colexes are much more 'rigid' than more general complexes.

\section{The model in $D$-manifolds}
\label{seccionIV}

\subsection{System and Hamiltonian}
\label{seccionIVA}

We now associate a physical system to a $D$-colex structure in a
$D$-manifold, $D\geq 2$. To this end, we place at each vertex (site)
a spin-$\frac 1 2$ system. To each $n$-cell $c$ we can attach the
cell operators
\begin{equation}\label{operador_ncelda}
B_c^{\sigma} := \bigotimes_{i\in I_c} \sigma_i,\qquad \sigma = X,Z
\end{equation}
where $X_i$ and $Z_i$ are the Pauli $\sigma_1$ and $\sigma_3$
matrices acting on the spin in the vertex $i$ and $I_c$ is the set
of vertices lying on the cell $c$. In order to generalize the
Hamiltonian \eqref{Hamiltoniano} we need sets of cells such that
their $X$ and $Z$ operators commute. But we have the following
result. For every $n$-cell $c_n$ and $m$-cell $c_m$ with $n+m>D+1$
\begin{equation}\label{ncommutacion}
\qquad [B_{c_n}^X,B_{c_m}^Z]=0.
\end{equation}
This is a consequence of the fact that $c_n$ and $c_m$ have colexes
with at least one color in common, because they have respectively
$p+1$ and $(q+1)$ colors. Then their intersection is a colex of
valence at least 1, and thus contains an even number of sites.

From this point on we choose fixed integers
$p,q\in\sset{1,\dots,D-1}$ with
\begin{equation}\label{}
p + q = D.
\end{equation}
The Hamiltonians that we propose is
\begin{equation}\label{nHamiltoniano}
H_{p,q}= - \sum_{c\in C_{p+1}} B_c^Z - \sum_{c\in C_{q+1}} B_c^X
\end{equation}
Again, color plays no role in the Hamiltonian. It is an exactly
solvable system and the ground state corresponds to a quantum error
correcting code with stabilizer the set of cell
operators\cite{gottesman}. We give a detailed calculation of the
degeneracy in appendix \ref{apendice_contaje}. The degeneracy is
$2^k$ with
\begin{equation}\label{ndegeneracion}
k=\binom D p h_p = \binom D {q} h_q,
\end{equation}
where  $h_p=h_q$  is the $p$-th Betty number
of the manifold. The ground states $\ket \psi$ are characterized by
the conditions
\begin{alignat}{2}
\paratodo c\in C_{p+1}\quad\quad &B_c^Z \ket \psi &= \ket \psi,\label{condicionesPZ}\\
\paratodo c\in C_{q+1} \quad\quad &B_c^X \ket \psi &= \ket
\psi.\label{condicionesQX}
\end{alignat}
Those eigenstates $\ket {\psi'}$ for which some of these conditions
are violated are excited states.  As in the $D=3$ case, excitations
have a local nature and we have a gapped system.

As a new feature respect to the three dimensional case, for $D\geq
4$ different combinations of the parameters $(p,q)$ are possible.
Each of these combinations gives rise to different topological
orders, thus making possible transitions between them. For example,
in $D=4$ the Hamiltonian
\begin{equation}\label{topologychanging}
H=H_{1,3}+\lambda H_{2,2}
\end{equation}
exhibits a topological phase transition as $\lambda$ is varied.

\subsection{Branes}
\label{seccionIVB}

In analogy with the string and membranes that appeared in the $D=3$
case, here we have to consider $p$-branes. For a $p$-brane we mean
an embedded $p$-manifold, closed or with a boundary. A $p$-brane is
homologous to zero when it is the boundary of a $p+1$-brane. Then
two $p$-branes are homologous if the $p$-brane obtained by their
combination is homologous to zero.

Let $Q$ be as usual the set of colors of the $D$-colex. Then for any
nonempty set $R\subsetneq Q$, a $R$-brane is a collection of
$R$-cells. It can be truly visualized as a $|R|$-brane in the
$R$-shrunk complex. There we see also that its boundary corresponds
to $\bar R$ cells. Let $b$ be a $R$-brane and $C_b$ its set of
$R$-cells. Then we can attach to $b$ operators $B_b^{\sigma} :=
\prod_{c\in C_b} B_c^\sigma$ for $\sigma = X,Z$. Suppose in
particular that $|R|$ = p and let $b$ be a closed $R$-brane. Then
$B_b^Z$ commutes with the Hamiltonian. If this were not the case,
then it would exist a $(q+1)$-cell, in particular an $\bar R$-cell
$c$, such that $\{B_b^Z,B_c^X\}=0$. But in that case, in the
$R$-shrunk complex the $p$-brane would have a boundary at the cell
coming from $c$. Similarly, closed $q$-brane $X$ operators also
commute with the Hamiltonian.

The operator $B_c^Z$ of a $(p+1)$-cell $c$ with color set $R\cup
\sset r$, $r\in Q-R$, is a closed $R$-brane. As the $R$-shrunk
complex reveals, it corresponds to the boundary of $c$. $B_c^Z$ acts
trivially in ground states \eqref{condicionesPZ}, and the same holds
true for any closed $p$-brane homologous to zero since it is a
combination of such operators $B_c^Z$. This is not the case for
closed $p$-branes which are not homologous to zero, and thus they
act nontrivially in the ground state.

\subsubsection{Equivalent branes}
It is natural to introduce an equivalence among those operators of
the form $\bigotimes_{v\in V} Z_v^{i_v}$, where $V$ is the set of
vertices of the colex and $i_v\in \sset{0,1}$. We say that two such
operators $O_1$ and $O_2$ are equivalent, $O_1\sim O_2$, if $O_1
O_2$ is a combination of $(p+1)$-cell operators $B_c^Z$. This
induces an equivalence among $p$-branes, since they have such an
operator attached. In fact, two $R$-branes $b$ and $b'$, with
$|R|=p$, are equivalent if and only if they are homologous. Observe
that two equivalent $p$-brane $Z$ operators produce the same result
when applied to a ground state. This motivates the introduction of
the notation $P_\mu^{R}$, $|R|=p$, for any operator $B_b^Z$ with $b$
a $R$-brane with homology labeled by $\mu$.

Likewise, we can introduce an equivalence among those operators
 of the form $\bigotimes_{v\in V}
X_v^{i_v}$ just as we have done for $Z$ operators. This induces an
equivalence relation among $q$-branes, and we use the notation
$Q_\nu^R$, $|R|=q$, for any $q$-brane operator $B_b^X$ with $b$ a
$R$-brane with homology labeled by $\nu$.

In appendix \ref{apendice_combinacion} we show that for any color
set $R\subset Q$ with $|R|=p-1$
\begin{equation}\label{}\label{combina_pbranas}
\prod_{r\in Q-R} P_\mu^{rR} \sim 0,
\end{equation}
where $rR$ is a shorthand for $\sset r \cup R$. Similarly if
$|R|=q-1$
\begin{equation}\label{}
\prod_{r\in Q-R} Q_\mu^{rR} \sim 0.
\end{equation}
These relations generalize \eqref{combina_cuerdas} and
\eqref{combina_membranas}. They give the interplay between homology
and color, and show that for each homology class only $\binom D p$
color combinations are independent, those which can be formed
without using one of the $D+1$ colors. This is why a combinatorial
number appears in the degeneracy of the ground state. The other
factor, $h_p$, follows from the fact that a homology basis for
$p$-branes has $h_p$ elements. Using the theory of quantum
stabilizer codes\cite{gottesman} one can see that by selecting a
basis for $p$-branes with labels $\mu=1,\dots,h_p$ and a color $r$
we can form a complete set of observables $\sset{P_\mu^R}_{\mu,
R\not\ni r}$.

\subsubsection{Commutation rules}

In general for suitable color sets $R,S$ we have
\begin{equation}\label{branas_commutan}
R\cap S\neq \emptyset\quad \Rightarrow \quad [P_\mu^R,Q_\nu^S]=0.
\end{equation}
This follows from the same reasoning used in \eqref{ncommutacion}.
We now explore the situation when $R$ and $S$ have no color in
common. Consider a basis $\sset{p_\mu}$ for closed $p$-branes.
Consider also a basis for $q$-branes $\sset{q_\nu}$, but chosen so
that $p_\mu$ and $q_\nu$ cross once if $\mu = \nu$, and do not cross
in other case. Then
\begin{equation}\label{nanticommutan}
R\cap S= \emptyset\quad \Rightarrow \quad P_\mu^R
Q_\nu^S=(-1)^{\delta_{\mu,\nu}} Q_\mu^S P_\nu^R.
\end{equation}
This can be reasoned without resorting to the geometrical picture.
Suppose that $[P_\mu^R,Q_\mu^S]=0$ and let $R=R'\cup\sset{r}$,
$Q-R-S=\sset{q}$. From \eqref{combina_pbranas} we have
$[\prod_{r'}P_\mu^{R'r'},Q_\mu^S]=0$. Then \eqref{branas_commutan}
implies $[P_\mu^{Rq},Q_\mu^S]=0$, and thus we have a homologically
nontrivial $q$-brane $X$ operator that commutes with all the
$p$-brane $Z$ operators. This being impossible, the assumption is
necessarily false.

\subsection{Excitations}
\label{seccionIVC}

There are two kinds of excitations, depending on wether a
$(p+1)$-cell or a $(q+1)$-cell condition is violated. We label
excitations with the color set of the cell they live in. Although we
focus on violation of $(q+1)$-cells, the situation is analogous for
$(p+1)$-cells.

Let $\ket \psi$ be a ground state and $b$ a $R$-brane, $R\subset Q$,
$|R|=p$. We first observe that $b$ has a boundary in the $R$-shrunk
complex at the cell corresponding to the $\bar R$-cell $c$ iff
\begin{equation}\label{condicion_excitacion}
\sset{P_b^R,B_c^X}=0.
\end{equation}
 But $B_b^Z\ket \psi$ has $\bar R$-excitation exactly at those
cells fulfilling \eqref{condicion_excitacion}. These means that the
excitation produced by the $p$-brane $b$ has the form of a
$p-1$-brane, precisely the boundary of $b$, $\partial b$.

Consider the particular case $p=1$. The excitations living at
$D$-cells are, as in the $D=3$ case, quasiparticles (anyons) with
color charge. In a connected manifold, we have $D$ constraints
generalizing \eqref{condicionesX}. They have the form
\begin{equation}\label{condiciones_particulas}
\prod_{c\in C_R} B_c^{X} = \prod_{c\in C_S} B_c^{X},
\end{equation}
where $|R|=|S|=D$ and $C_R$ is the set of $R$-cells. These relations
imply that the number of particles of each color must agree in their
parity. Therefore, from the vacuum we can create pairs of particles
of a single color or groups of $D+1$ particles, one of each color.
This is completely anologous to $D=3$.

Now suppose that $p>1$. We have seen that excitations can be created
as the boundary of a $p$-brane. If in particular it is a $R$-brane,
then excitations live in $\bar R$-cells. It is natural an
interpretation of these excitations as some kind of
$(p-1)$-dimensional flux, a $\bar R$-branyon. Then it must be
conserved, and in fact for each $(q+2)$-cell $c$ we have the
constraints:
\begin{equation}\label{condicionesCelda}
\prod_{c\in C_R^c} B_c^X = \prod_{c\in C_S^c} B_c^X,
\end{equation}
where $|R|=|S|=q+1$ and $C_R^c$ is the set of $R$-cells lying on the
cell $c$. This is a generalization of \eqref{condicionesZ} and is in
agreement with \eqref{combina_pbranas}.

Finally, as in the 3-dimensional case, we can wind branyons around
each other and sometimes get a global phase. Let $|R|=q+1$ and
$|S|=p+1$. Then when a $R$-branyon winds around a $S$-branyon, the
system gets a global minus sign iff $|R\cap S|=1$, as follows from
the commutation rules \eqref{nanticommutan}.

\section{Conclusions}
\label{seccionV}

In this paper we have explored topological orders in $D=3$
by means of models for  quantum lattice Hamiltonians constructed
with spins $S=\half$ located at lattice sites.
These models are exactly solvable and this is a feature that allows
us to explore the quantum properties of the whole spectrum. The ground
state is found to be in a string-net condensate, or alternatively in a
membrane-net condensate. This type of membrane-net condensation is an
interesting feature of our models that do not appear in 3D toric codes.
In dimensions higher than $D=3$ we have also extended the construction
of our models and found brane-net condensation. As for excitations,
they are either quasiparticles or certain type of extended fluxes.
These excitation show unusual braiding statistics properties similar
to anyons in $D=2$, and we call them branyons since they involve extended
objects associated to branes.

Another interesting result is the possibility of having a topology-changing
transition between two topologically ordered phases that we find
with our models in $D=4$.
We may wonder whether it is possible to have a similar topology-changing
process in dimension $D=3$ as in \eqref{topologychanging}.
One obvious way to achieve this is by
using the construction in $D=4$ and flatten it into $D=3$, thereby
reducing the dimensionality of the interaction but at the expense
of loosing the locality of the interaction.

There does not exist a fully or complete topological order in $D=3$
dimensions, unlike in $D=2$. That is to say, there does not exist a
topological order that can discriminate among all the possible
topologies in three dimensional manifolds. We may introduce the
notion of a Topologically Complete (TC) class of quantum
Hamiltonians when they have the property that their ground state
degeneracy (and similarly for excitations) is different depending on
the topology of the manifold where the lattice is defined on. From
this perspective, we have found a class of topological orders based
on the construction of certain lattices called colexes that can
distinguish between 3D-manifolds with different homology properties.
Homology is a topological invariant, but not enough to account for
the whole set of topologically inequivalent manifolds in $D=3$. For
instance, the famous Poincar\'e sphere is an example of 3D-manifold
that has the same homology as a 3-sphere. Poincar\'e was able to
proof that the fundamental group (or first homotopy group) of this
new sphere has order 120. As the standard 3-sphere has trivial
fundamental group, they are different. Since then, many other
examples of homology spheres have been constructed that are
different topological structures. In this regard, we could envisage
the possibility of finding a quantum lattice Hamiltonian, possibly
with a non-abelian lattice gauge theory, that could distinguish
between any topology in three dimensions by means of its ground
state degeneracy. This would amount to solving the Poincar\'e
conjecture with quantum mechanics.

From the viewpoint of quantum information, the
topologically ordered ground states that we have constructed provide
us with an example of topological quantum memory: a reservoir of
states that are intrinsically robust against decoherence due to
the encoding of information in the topology of the system.

 \noindent {\em Acknowledgements} We acknowledge financial
support from a PFI fellowship of the EJ-GV (H.B.), DGS grant  under
contract BFM 2003-05316-C02-01 and INSTANS (M.A.MD.), and CAM-UCM grant under
ref. 910758.

\appendix

\section{Connected sum}\label{apendice_suma}

This appendix is not strictly necessary in the logical structure of
the text, but we include it because it gives a beautiful example of
how the topological structure of a colex is contained in its
c-graph.

Given two connected $D$-manifolds $\mathcal M_1$ and $\mathcal M_2$,
their connected sum is constructed as follows. First choose a pair
of $D$-discs, one in each manifold, and delete its interior. Then we
glue together their boundaries. The resulting object is a
$D$-manifold, denoted $\mathcal M_1\sharp \mathcal M_2$.

As we have already noted, a c-graph contains all the information
about its manifold, if it yields any. Thus one expects that the
connected sum of two manifolds can also be expressed in terms of
c-graph manipulations. In fact, this is true, as we explain now.

Consider two $D$-colexes, $\mathcal C_1$ and $\mathcal C_2$. In
order to construct $\mathcal C_1\sharp\mathcal C_2$, we choose a
vertex of each of the colexes and select small neighborhoods of
these vertices. Due to the one-to-one correspondence between
c-graphs and cells, this two neighborhoods of vertices are identical
at a complex and color level. Thus the gluing operation between
boundaries can be chosen so that the structure of the complex is
preserved and the coloring of edges coincides. Therefore, the result
of the operation is a $D$-colex.

\begin{figure}
\includegraphics[width=8 cm]{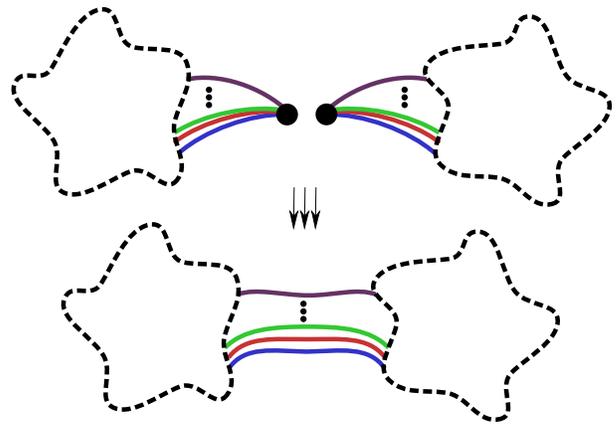}
\caption{The connected sum of two colexes can be performed at a
$c$-graph level. Two vertexes must be selected and deleted. Then the
edges that have been cut must be joined according to their
color.}\label{figura_suma_conexa}
\end{figure}

How do we express the process at a c-graph level? Simply choose one
vertex at each c-graph, and delete them. Then connect the edges
according to their color. Fig.~\ref{figura_suma_conexa} displays the
procedure.

\section{How to construct D-colexes}\label{apendice_inflado}

\begin{figure}
\includegraphics[width=8.5 cm]{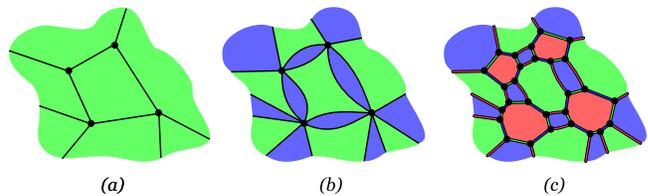}
\caption{This figure explains a process that converts an arbitrary
2-complex on a surface into a 2-colex. In (a) green color is given
to all the 2-cells of the 2-complex. In (b) 1-cells are inflated to
give blue 2-cells. Finally in (c) 0-cells are inflated to give red
2-cells and 1-cells are accordingly colored.}\label{figura_inflado}
\end{figure}

We present a procedure to construct colexes in arbitrary closed
manifolds. The idea is to start with an arbitrary complex and
inflate its cells till a colex is obtained. We now explain the
process in detail. It is illustrated in Fig.~\ref{figura_inflado}.

First we have to state what we mean by inflating a $n$-cell, $0\leq
n\leq D$. The idea is to preserve the boundaries of the cell
untouched but inflate all other points in order to obtain a
$D$-cell. For each $(n+l)$-cell that belongs to the boundary of the
inflated cell, we must introduce a $(n+l-1)$-cell. The inflation of
cells of the same dimension can be performed in any order, and all
the cells must be inflated. Inflation starts with $(n-1)$-cells,
then continues with $(n-2)$-cells, and so on, till 0-cells are
inflated in the end.

We can proof that this procedure gives a $D$-colex by an inductive
argument. We will need some facts. First, we observe that the
$D$-cells of a $D$-colex can be labeled with the unique color which
its subcolex does not contain. Conversely, if we can color the
$D$-cells of a $D$-colex with $D+1$ colors in such a way that
neighboring cells have different colors, then we can color edges
according to the $D$-cells they connect. Note also that for each
cell in the original $D$-complex, the inflated one has a $D$-cell.
This means that we can label inflated $D$-cells with the dimension
of the cell in the original complex.

Finally we proceed with the proof. The case $D=1$ is trivial. We
suppose that the procedure works for $(D)$-manifolds, and check it
for $(D+1)$-manifolds. To this end, consider the boundary of any
inflated $(D+1)$-cell which comes from the inflation of a 0-cell.
Imagine all the inflation process projected into this (fixed)
$D$-sphere. In the beginning, we can see a complex in this sphere.
Its vertices correspond to edges that cross the surface, edges to
faces that cross it, and so on. As the inflation proceeds in the
original complex, the projected complex is also inflated. When
1-cells are inflated, the projected complex has become a $D$-colex
because of the induction hypothesis. Thus it can be properly
colored. Moreover, we can perform the coloring on its $D$-cells
using the labels
 attached to the corresponding $(D+1)$-cells in the inflated $(D+1)$-complex.
 From this coloring we deduce a coloring for the edges of the $D$-colex.
 In fact, all this is true for each of the subgraphs in the surfaces of
 $(D+1)$-cells obtained by inflation of 0-cells. Finally, we give a different
 color to the edges that are not contained in these surfaces. Checking that this
 coloring gives the desired properties that make the complex a colex is now easy.

\section{Brane combination}\label{apendice_combinacion}

Consider a $D$-colex with color set $Q$. Let
 $b_R$ be a closed $R$-brane, $\emptyset\subsetneq R\subsetneq Q$. It
is the purpose of this section to show that for any $r\in R$ there
exist a family of closed $|R|$-branes $b_S$ homologous to $B_R$ such
that
\begin{equation}\label{combinacion_directa}
B_{b_R}^\sigma = \prod_S B_{b_S}^\sigma.
\end{equation}
The sum extends over all $S\subset \bar r:=Q-\sset{r}$ with
$|S|=|R|$.

We first consider the case $R=\sset{r}$. Then $b_R$ is a string. It
consists of $r$-egdes that link $\bar R$-cells. $B_{b_R}^\sigma$
acts nontrivially in an even number of vertices per $\bar R$-cell.
Thus we can gather them together in pairs, and connect them through
a path which only contains edges with colors in $Q-R$. Then for each
$s\in Q- R$, the set of all $s$-edges we have used forms a string
$b_S$, $S={s}$. Then, certainly \eqref{combinacion_directa} holds
true and each string $b_S$ is closed because the r.h.s. commutes
with operators from $\bar S$ cells and so the l.h.s. also does.

Now consider the case $|R|>1$. Let $\tilde r:=R-\sset r$. Consider
the restriction of $B_{b_R}^\sigma$ to any $\bar r$-cell $c$,
denoted $B_{b}^\sigma$. This operator corresponds to a closed
$\tilde r$-brane $b$ in the $(D-1)$-colex that forms the boundary of
$c$. Since this colex is a sphere, $b$ is a boundary and thus
$B_b^\sigma$ is a combination of $|R|$-cells. As we did for strings,
we can do this for every $\bar r$-cell, gather cells together by
color and form the required closed $|R|$-branes.

\section{Degeneracy of the ground state}\label{apendice_contaje}

In the theory of quantum error correcting codes, the ground state of
the Hamiltonian \eqref{nHamiltoniano} is called a stabilizer code
\cite{gottesman}. Thus, the theory of stabilizer codes naturally
fits to study the degeneracy, but we will avoid to use its language
although this makes the exposition less direct.

The ground state of the Hamiltonian \eqref{nHamiltoniano} is the
intersection of subspaces of eigenvalue 1 of $(p+1)$-cell and
$(q+1)$-cell operators, as expressed in equations
\eqref{condicionesPZ} and \eqref{condicionesQX}. This subspace has
an associated projector, which in turn will be the product of the
projectors onto each of the subspaces of eigenvalue 1:
\begin{equation}\label{projector}
\prod_{c\in C_{p+1}} \frac 1 2 (1+B_c^Z) \prod_{c\in C_{q+1}} \frac
1 2 (1+B_c^X).
\end{equation}
Each of these projectors reduces the dimension of the space by a
half, but not all of them are independent, because certain relations
among cell operators exist. For $(p+1)$-cells these relations have
the form
\begin{equation}\label{relacion_tipoZ}
\prod_{c\in C_{p+1}} (B_c^Z)^{i_c}=1,
\end{equation}
where $i_c=0,1$. Analogous relations exist for $(q+1)$-cells:
\begin{equation}\label{relacion_tipoX}
\prod_{c\in C_{q+1}} (B_c^X)^{i_c}=1.
\end{equation}
If the number of spins is $n$ and the number of independent
projectors is $l$, then the degeneracy of the ground state will be
$2^k$ with $k=n-l$. Suppose that the number of independent relations
of type \eqref{relacion_tipoZ} is $l_1$ and that for relations
\eqref{relacion_tipoX} this number is $l_2$. Then we have
$l=|C_{p+1}|-l_1+|C_{q+1}|-l_2$. Our starting point is then the
equation
\begin{equation}\label{basica_degeneracion}
k=|C_0|-|C_{p+1}|-|C_{q+1}|+I(D,p+1)+I(D,q+1),
\end{equation}
where $n=|C_0|$ is the number of sites and $I(D,s)$ is the number of
independent relations among $s$-cells in a $D$-colex.

$I(D,s)$ only depends upon the cardinalities of cell sets $|C_i|$
and the Betty numbers of the manifold $h_i$, as we will show
calculating its value recursively. First, we note that
\begin{equation}\label{}
I(D,D)=dh_0,
\end{equation}
because the unique independent relations in this case are those in
\eqref{condiciones_particulas}, for each connected component. For
$s<D$, a relation between cells has the general form
\begin{equation}\label{relacion_generica}
\prod_{|S|=s} \prod_{c\in D_S} B_c^\sigma = 1,
\end{equation}
where $D_S\subset C_S$. Let $r\in Q$ be a color. If we only consider
those relations that include color sets $R\subset \bar r$ we
effectively reduce the problem by 1 dimension. Gathering together
all such relations that appear in $\bar r$-cells we get a total
count of
\begin{equation}\label{}
I_{\bar r}(D,s)=I(D-1,s)|_{h_0=h_D=|C_{\bar r}|, h_{i\neq 0,D}=0}.
\end{equation}
Since the r.h.s. of \eqref{relacion_generica} commutes with any cell
operator, in fact the relation has the form
\begin{equation}\label{relacion_branas}
\prod_{|S|=s} B_{b_S}^\sigma = 1,
\end{equation}
where $b_S$ is a \emph{closed} $S$-branes $b_S$. Then consider any
such relation in which a $R$-brane $b_R$ appears with $r\in R$. If
we have at hand all the relations of the form
\eqref{combinacion_directa}, we can use them to eliminate the term
$b_R$ in \eqref{relacion_branas}. This can be done for every such
$R$, till a relation containing only colors in $\bar r$ is obtained.
Therefore, our following task is to count how many of the relations
\eqref{combinacion_directa} are independent for each $R$.

Suppose then that we have a relation of the form
\eqref{combinacion_directa} that follows from other $t$ relations of
the same form (but not from a subset of them):
\begin{equation}\label{combinacion_directa_indice}
B_{b_{R,i}}^\sigma = \prod_S B_{b_{S,i}}^\sigma\quad\quad
i=1,\dots,t.
\end{equation}
Then for the l.h.s of the relations the following is true
\begin{equation}\label{combinacion_indices}
B_{b_R}=\prod_{i=1}^t B_{b_{R,i}}.
\end{equation}
Since all the branes that appear in \eqref{combinacion_directa} are
$R$-branes, the equation can be interpreted in terms of $Z_2$-chains
of $|R|$-cells in the $R$-shrunk complex. It states that
$b_R=b_{R,1}+\dots+b_{R,t}$. The argument can be reversed; any such
a dependence between $|R|$-cycles in the $R$-shrunk complex
corresponds to a dependence among relations of the form
\eqref{combinacion_directa}.

Therefore, counting the number of independent relations of the form
\eqref{combinacion_directa} for a given $R$ amounts to count the
number of independent $\Z_2$-chains of closed $|R|$-cycles in the
$R$-shrunk complex. For $|R|=s$, this number is
\begin{equation}\label{}
S(D,s)= \sum_{i=0}^{n-s} (-1)^i h_{s+i} + \sum_{i=1}^{n-s} (-1)^i
|C_{s+i}|.
\end{equation}
This follows by recursion. $S(D,D)=h_0$ and
$S(D,s)=h_{D-s}+(|C_{s+1}|-S(D,s+1))$.

We have to consider all the possible sets $R$ in which $r$ is
contained:
\begin{equation}\label{}
I_r(D,s) = \sum_{R\ni r \atop |R|=s} S(D,s)|_{\text{$R$-shrunk}}.
\end{equation}
Then
\begin{equation}\label{}
I(D,s) = I_{\bar r}(D,s)+I_r(D,s),
\end{equation}
which can be solved and
gives
\begin{multline}\label{cuenta_independientes}
I(D,s) = \binom D {s-1} \sum_{i=0}^{D-s} (-1)^i h_{s+i}+\\
+\sum_{i=0}^{D-s-1} \binom {s+i} {s-1} (-1)^i |C_{s+i+1}|
\end{multline}
Now recall equations \eqref{Euler_plus}. We can sum up these
equations for $r=0,\dots,s$ with an alternating sign $(-1)^r$. Using
the fact that
\begin{equation}\label{}
\sum_{i=0}^a \binom {b+1} i (-1)^i = (-1)^a\binom b a
\end{equation}
we get
\begin{multline}\label{combina_euler}
\binom D s \chi = (-1)^s C_0 + \sum_{i=0}^{s-1}\binom {D-i-1}{D-s}
(-1)^i|C_{D-i}|+
\\+\sum_{i=r+1}^{D}\binom {i-1}{s} (-1)^i
|C_i|.
\end{multline}
Gathering together \eqref{basica_degeneracion},
\eqref{cuenta_independientes} and \eqref{combina_euler} we finally
obtain \eqref{ndegeneracion}.


\begin{thebibliography}{99}

\bibitem{landau37}
L. D. Landau, Phys. Z. Sowjetunion {\bf 11}, 26 (1937).

\bibitem{ginzburg_landau50}
V. L. Ginzburg, L. D. Landau, Zh. Ekaper. Teoret. Fiz. {\bf 20},
1064 (1950).

\bibitem{wenbook04}
X.-G. Wen. {\em Quantum Field Theory of Many-body Systems}, Oxford
University Press, (2004).

\bibitem{wenniu90}
X.-G. Wen and Q. Niu, Phys. Rev. {\bf B 41}, 9377 (1990).

\bibitem{wen90}
X.-G. Wen, Int. J. Mod. Phys. {\bf B 4}, 239 (1990).

\bibitem{wen92}
X.-G. Wen, Int. J. Mod. Phys. {\bf B 6}, 1711 (1992).


\bibitem{kitaevpreskill06}
A. Kitaev and J. Preskill,
 Phys. Rev. Lett. {\bf 96}, 110404 (2006)

\bibitem{levinwen06}
M. Levin and X.-G. Wen, Phys. Rev. Lett. {\bf 96}, 110405 (2006).

\bibitem{blokwen90}
X.-G. Wen, Phys. Rev. {\bf B 42}, 8145 (1990).

\bibitem{read90}
N. Read, Phys. Rev. Lett. {\bf 65}, 1502 (1990).

\bibitem{frohlichkerler91}
J. Fro\"{o}hlich and T. Kerler, Nucl. Phys. {\bf B 354}, 369 (1991).

\bibitem{roksharkivelson88}
D.S. Rokshar and S.A. Kivelson,
Phys. Rev. Lett. {\bf 61}, 2376 (1988).

\bibitem{readchakraborty89}
N. Read and B. Chakraborty,
Phys. Rev. {\bf B 40}, 7133 (1989).

\bibitem{moessnersondhi01}
R. Moessner and S.L. Sondhi,
Phys. Rev. Lett. {\bf 86}, 1881 (2001).

\bibitem{ardonne_etal04}
E. Ardonne, P. Fendley and E. Fradkin,
Annals  of Phys. {\bf 310}, 493 (2004).


\bibitem{kalmeyerlaughlin87}
V. Kalmeyer and R.B. Laughlin,
Phys. Rev. Lett. {\bf 59}, 2095 (1987).

\bibitem{wenwilczekzee89}
X. G. Wen, F. Wilczek, and A. Zee,
Phys. Rev. B {\bf 39}, 11413 (1989).

\bibitem{readsachdev91}
N. Read and S. Sachdev,
Phys. Rev. Lett. {\bf 66}, 1773 (1991).

\bibitem{wen91}
X.-G. Wen,
Phys. Rev. B {\bf 44}, 2664 (1991).


\bibitem{senthilfisher00}
Phys. Rev. Lett. {\bf  86}, 292 (2001).

\bibitem{wen02}
X.-G. Wen,
Phys. Rev. B {\bf 65}, 165113 (2002).


\bibitem{sachdevparks02}
S. Sachdev and K. Park,
Annals  of Phys. {\bf 298}, 58 (2002).



\bibitem{balentsfishergirvin02}
L. Balents, M. P. A. Fisher, and S. M. Girvin
Phys. Rev. B {\bf 65}, 224412 (2002).


\bibitem{martindelgado04}
F. Verstraete, M.A. Martin-Delgado, J.I. Cirac, Phys. Rev.
Lett. {\bf 92}, 087201 (2004).


\bibitem{martindelgado04b}
J. J. Garcia-Ripoll, M. A. Martin-Delgado, J. I. Cirac, Phys.
Rev. Lett. {\bf 93}, 250405 (2004).

\bibitem{duandemlerlukin03} L.-M.\,Duan, E.\,Demler, M.\,D.\,Lukin,
 Phys.\ Rev.\ Lett.\ \textbf{91}, 090402 (2003),
\texttt{cond-mat/0210564}.

\bibitem{zoller05}
 A. Micheli, G.K. Brennen, P. Zoller,
quant-ph/0512222.


\bibitem{pachos05}
J. K. Pachos,  quant-ph/0511273.

\bibitem{kitaev05}
A. Kitaev,
cond-mat/0506438.

\bibitem{rmp}
A. Galindo and M.A. Martin-Delgado,
Rev.Mod.Phys. {\bf 74}  347 (2002); quant-ph/0112105.


\bibitem{kitaev97}
A.\,Yu.\,Kitaev,
Annals of Physics \textbf{303} no.~1, 2--30 (2003), \texttt{quant-ph/9707021}.

\bibitem{freedman98}
M. H. Freedman,
Proc. Natl. Acad. Sci., USA, {\bf 95} 98-101, (1998).

\bibitem{dennis_etal02}
E. Dennis, A. Kitaev, A. Landahl, J. Preskill,
J. Math. Phys. {\bf 43}, 4452-4505 (2002).

\bibitem{bravyikitaev98}
 S. B. Bravyi, A. Yu. Kitaev,
quant-ph/9811052.

\bibitem{Ogburn99}
R. W. Ogburn and J. Preskill,
Lecture Notes in Computer Science {\bf 1509},
341--356, (1999).

\bibitem{freedman_etal00a}
Michael H. Freedman, Alexei Kitaev, Zhenghan Wang,
Commun.Math.Phys. {\bf 227}  587-603, (2002).

\bibitem{freedman_etal00b}
M. Freedman, M. Larsen, Z. Wang,
Comm.Math. Phys. {\bf 227}  605--622, (2002).

\bibitem{freedman_etal01}
M. H. Freedman, A. Kitaev, M. J. Larsen, Z. Wang,
Bull. Amer. Math. Soc. {\bf 40}   31-38, (2003);
quant-ph/0101025.

\bibitem{preskillnotes} J. Preskill,
{\em Lecture notes on Topological  Quantum \linebreak Computation},
http://www.theory.caltech.edu/preskill/\linebreak ph219/topological.ps.

\bibitem{topologicalclifford}
H. Bombin and M.A. Martin-Delgado; quant-ph/0605138.

\bibitem{homologicalerror}
H. Bombin and M.A. Martin-Delgado; quant-ph/0605094.


\bibitem{optimalGraphs}
 H. Bombin and M.A. Martin-Delgado,
 Phys. Rev. A {\bf 73}, 062303 (2006); quant-ph/0602063.


\bibitem{thurston}
W. Thurston,  ``Three-dimensional geometry and topology''.
Vol. 1. Edited by Silvio Levy. Princeton Mathematical Series, {\bf 35}.
Princeton University Press, Princeton, NJ, 1997

\bibitem{perelmanI}
G. Perelman, math.DG/0211159.

\bibitem{perelmanII}
G. Perelman, math.DG/0303109


\bibitem{perelmanIII}
G. Perelman, math.DG/0307245

\bibitem{levinwen05}
M. Levin and X.-G. Wen,  Phys. Rev. {\bf B 71}, 045110
(2005).

\bibitem{hammazanardiwen05}
A. Hamma, P. Zanardi, X.-G. Wen,
Phys.Rev. B {\bf 72},  035307 (2005).

\bibitem{wang_etal03}
C. Wang, J. Harrington, J. Preskill,  Annals Phys. {\bf 303}  31-58, (2003).


\bibitem{takeda04}
K. Takeda, H. Nishimori, Nucl.  Phys. B {\bf 686} 377, (2004)

\bibitem{wilczek82}
F. Wilczek,
 Phys. Rev. Lett. {\bf 49}, 957 (1982)

\bibitem{leinaasmyrheim77}
J.M. Leinaas and J. Myrheim, Nuo. Cim. B {\bf 37}, 1 (1977).

 \bibitem{gottesman}
 D. Gottesman,
 Phys. Rev. A {\bf 54}, 1862, (1996).



\end{thebibliography}
\end{document}